# An Experimental-based Review of Image Enhancement and Image Restoration Methods for Underwater Imaging

**Yan Wang[1], Wei Song[2] (✉), Giancarlo Fortino[3], Lizhe Qi[1], Wenqiang Zhang[4], Antonio Liotta[5]**

[1] Academy for Engineering & Technology, Fudan University, China
[2] College of Information Technology, Shanghai Ocean University, Shanghai 201306, China
[3] University of Calabria, Italy
[4] Shanghai Key Laboratory of Intelligent Information Processing, School of Computer Science, Fudan University, Shanghai 200433, China
[5] Napier Edinburgh University, School of Computing, Edinburgh, U.K.

Corresponding author: Wei Song (e-mail: wsong@shou.edu.cn)

This work was supported by the National Natural Science Foundation of China (NSFC) Grant 61702323, the Program for Professor of Special Appointment (Eastern Scholar at Shanghai Institutions of Higher Learning No. TP2016038).

**ABSTRACT** Underwater images play a key role in ocean exploration, but often suffer from severe quality degradation due to light absorption and scattering in water medium. Although major breakthroughs have been made recently in the general area of image enhancement and restoration, the applicability of new methods for improving the quality of underwater images has not specifically been captured. In this paper, we review the image enhancement and restoration methods that tackle typical underwater image impairments, including some extreme degradations and distortions. Firstly, we introduce the key causes of quality reduction in underwater images, in terms of the underwater image formation model (IFM). Then, we review underwater restoration methods, considering both the IFM-free and the IFM-based approaches. Next, we present an experimental-based comparative evaluation of state-of-the-art IFM-free and IFM-based methods, considering also the prior-based parameter estimation algorithms of the IFM-based methods, using both subjective and objective analysis (the used code is freely available at https://github.com/wangyanckxx/Single-Underwater-Image-Enhancement-and-Color-Restoration). Starting from this study, we pinpoint the key shortcomings of existing methods, drawing recommendations for future research in this area. Our review of underwater image enhancement and restoration provides researchers with the necessary background to appreciate challenges and opportunities in this important field.

## I. INTRODUCTION

The oceans contain unknown creatures and vast energy resources, playing an important role in the continuation of life on earth [1]. Hence significant effort has been dedicated worldwide, since the middle of the 20th century, to actively engage in high-tech marine exploration activities. Vision technology has attracted significant attention, for its ability to carry high information density [2]. Researchers strive to capture high-quality underwater images for a variety of underwater applications, including robotics [3], rescue missions, man-made structures inspection, ecological monitoring, and real-time navigation [4], [5].

However, the quality of underwater images is severely affected by the particular physical and chemical characteristics of underwater conditions, raising issues that are more easily overcome in terrestrial imaging.

Underwater images always show color cast, e.g., green-bluish color, which is caused by different attenuation ratios of red, green and blue lights. Also, the particles that are suspended underwater absorb the majority of light energy and change the direction of light before the light reflected from underwater scene reaches the camera, which leads to images having low-contrast, blur and haze [6].

In order to increase the range of underwater imaging, artificial light sources are often. Yet, artificial light too is affected by absorption and scattering [7]. At the same time, non-uniform illumination is introduced, resulting in bright spots at the center of the underwater image, with insufficient illumination towards the boards. Other quality degradation phenomena include, for instance shadowing. Thus, extracting valuable information for underwater scenes requires effective methods to correct color, improve clarity and address blurring and background scattering, which is the aim of image enhancement and restoration algorithms. These are particularly challenging due to the complex underwater environment, where images are degraded by the influence of water turbidity, light absorption and scattering, which may change broadly.

Understanding the underwater optical imaging model could help us better design and propose robust and effective enhancement strategies. Fig. 1 shows the underwater optical imaging process and the selective attenuation of light, which is drawn and modified based on the model proposed by Huang et al. [8]. The selective attenuation characteristics is shown on the right side of Fig. 1. When travelling through water, the red light – having a longer wavelength – is absorbed faster than green and blue wavelengths (which are shorter). That is why underwater images often appear to have green-bluish tones.

Figure 1 shows the interaction between light, transmission medium, camera and scene. The camera receives three types of light energy in line of sight (LOS): the direct transmission light energy reflected from the scene captured (direct transmission); the light from the scene that is scattered by small particles but still reaches the camera (forward scattering); and the light coming from atmospheric light and reflected by the suspended particles (background scattering) [9]. In the real-world underwater scene, the use of artificial light sources tends to aggravate the adverse effect of background scattering. The particles suspended underwater generated unwanted noise and aggravate the visibility of dimming images. The imaging process of underwater images can also be represented as the linear superposition of the above three components [10], [11] and shown as follows:

$$E_T(x,y) = E_d(x,y) + E_f(x,y) + E_b(x,y) \quad (1)$$

Whereby $(x,y)$ represents the coordinates of individual image pixels; $E_T(x,y)$, $E_d(x,y)$, $E_f(x,y)$, and $E_b(x,y)$ represent the total signal energy captured by the camera, the direct transmission component, the forward scattering component, and the background scattering component, respectively. Since the distance between the underwater scene and the camera is relatively close, the forward scattering component can be ignored and only the direct transmission and background scattering components [12]–[16] are considered.

If we define $J$ as the underwater scene, $t$ as the residual energy ratio after $J$ was captured by the camera, $B$ as the homogenous background light, then the scene captured by the camera $I$ can be represented as in Eq.(2), which is considered as the simplified underwater image imaging model (IFM).

$$I^c(x) = J^c(x)t^c(x) + B^c(1 - t^c(x)) \quad (2)$$

Whereby $x$ represents one particular point $(i,j)$ of the scene image, $c$ is one of the red, green and blue (RGB) channels, and $J^c(x)t^c(x)$ and $B^c(1 - t^c(x))$ represent the direct transmission and background scattering component, respectively.

The visibility of underwater images can be improved using hardware [17]–[22] and software solutions [15], [23]–[26].

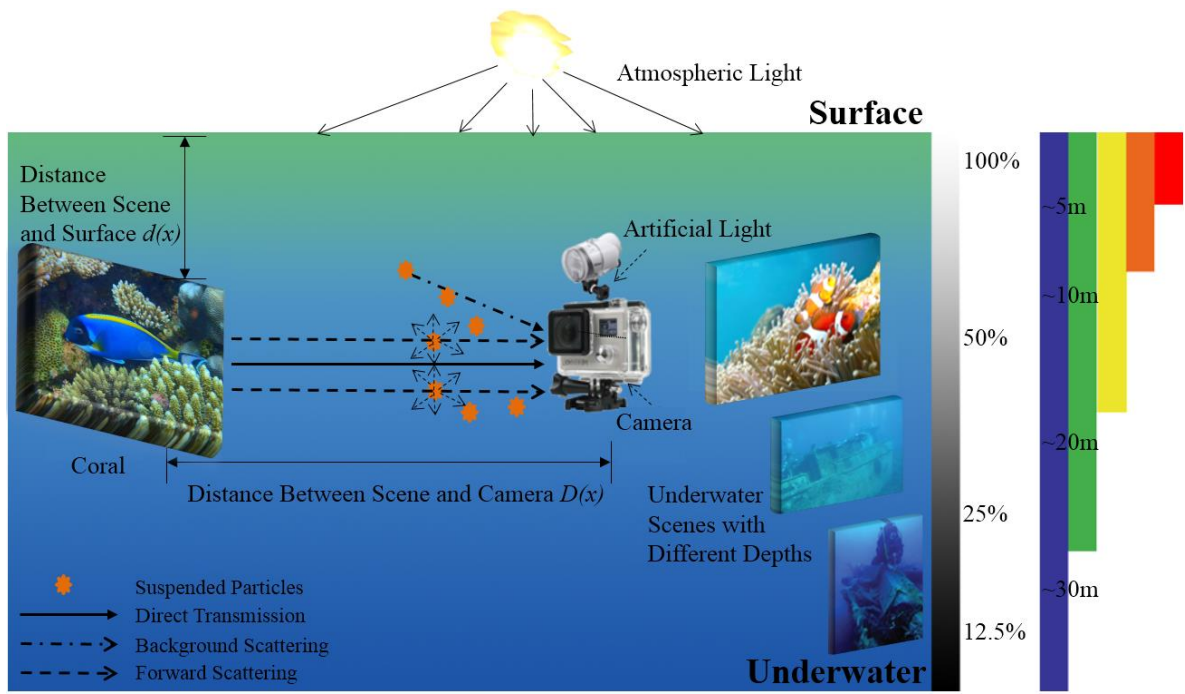

**FIGURE 1.** Diagram of underwater optical imaging.

The specialized hardware platforms and cameras can be expensive and power-consuming. What is more, they are not adaptive to different underwater environments. Thus, many algorithmic methods have been developed for underwater image quality improvement by image enhancement or restoration.

Although some reviews of underwater image enhancement and restoration have been published, these tend to only concentrate on certain aspects of underwater processing. For example, Kaeli et al. [27] focused on algorithms for underwater image color correction; Sahu et al. [28] introduced limited underwater image enhancement methods. Lu et al. [29] and Han et al. [30] reviewed more aspects of underwater optical processing, including underwater image de-scattering, underwater image restoration, underwater image quality assessments, and future trends and challenges in designing and processing underwater images.

Nonetheless, several issues are not fully addressed in previous reviews: 1) the existing classifications are incomplete, and miss the very latest developments based (for instance) on deep learning; 2) it remains unclear the extent by which specific methods lead to improving image quality and how.

In this paper, we address the above problems, providing a broader review, an experimental-based comparison of key methods, and providing an up-to-date snapshot of challenges and future directions.

It should be noted that we focus specifically on quality improvement algorithms for single underwater image. Our contributions to the study of quality improvement of underwater images are multi-fold:

(1) We categorize the quality improvement methods of underwater images into two broad classes: IFM-free image enhancement methods and IFM-based image restoration methods, from the perspective that these improve image quality either through the optical imaging physical model or not. The categories of the reviewed methods are shown in Fig 2, which will help better understanding which models are suited best for which problem domain.

(2) We carry out experimental-based comparisons of some representative algorithms from both IFM-free and IFM-based categories, providing an evaluation based on different quality metrics. For the sake of replicability, we have made all the code available at https://github.com/wangyanckxx/Single-Underwater-Image-Enhancement-and-Color-Restoration.

(3) We provide a critical evaluation of image restoration methods based on prior-knowledge, which reveals the issues that these raise in estimating the parameters of underwater image restoration.

(4) We share our lessons-learnt from working in this area, revealing latent difficulties and problems faced in underwater image quality improvement.

The rest of the paper is organized as follows. Section II, provides a review of IFM-free image enhancement methods, followed by an overview of IFM-based image restoration methods in Section III. The experimental-based comparisons of different methods for underwater image improvement is presented in Section IV. Finally, discussion and future work directions are stated in section V.

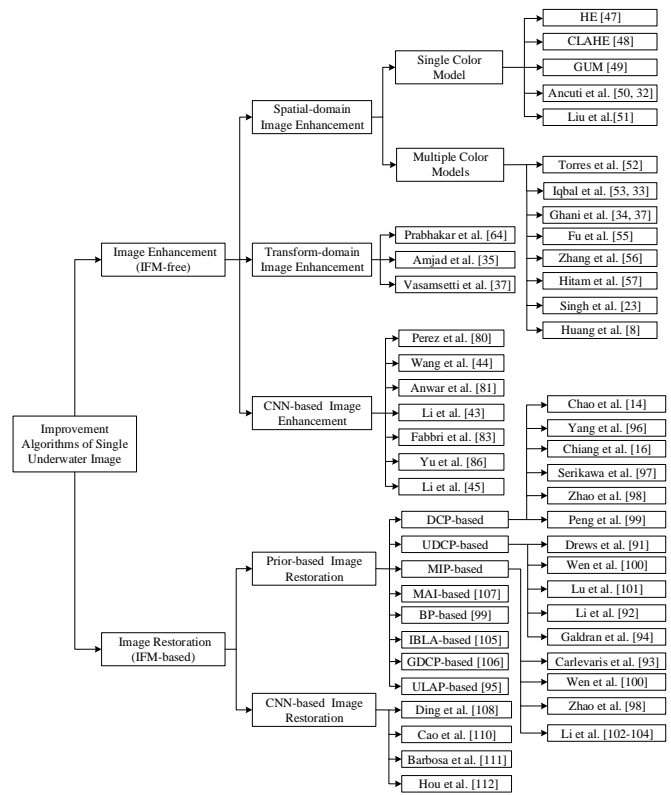

**FIGURE 2.** Categories of quality improvement of single underwater image.

## II. IFM-FREE IMAGE ENHANCEMENT

IFM-free underwater image enhancement methods improve the contrast and color of images mainly based on pixel intensity re-distribution, without considering the particular underwater imaging principles. Early studies of underwater image enhancement often apply outdoor image enhancement methods directly to underwater images. Later methods are specially designed according to the characteristics of the underwater image, e.g., hazing, color cast, and low contrast. These methods change the pixels values in either the spatial domain or a transformed domain. Recently, deep learning models, especially Convolutional Neural Networks (CNN), have been used for image enhancement, based on the idea that hidden features may be learned for quality enhancement.

In this review, we separate the IFM-free methods in three subclasses: spatial-domain image enhancement [31]–[37]; transform-domain image enhancement [38]–[42]; and CNN-based image enhancement [43]–[45].

## A. SPATIAL-DOMAIN IMAGE ENHANCEMENT

The histograms of underwater images show a relatively more concentrated distribution of pixel values than in natural images. Therefore, expanding the dynamic range of the image histogram provides a way for enhancing the visibility of underwater images. Spatial-domain image enhancement methods complete an intensity histogram redistribution by expanding gray levels based on the gray mapping theory [46]. This can be done in different color models. Common color models include Red-Green-Blue (RBG), Hue-Saturation-Intensity (HSI), Hue-Saturation-Value (HSV) and CIE-Lab. Based on whether single a color model (SCM) or multiple color model (MCM) is used in the histogram redistribution process, we can divide spatial-domain image enhancement methods into SCM-based and MCM-based image enhancement.

*1) SCM-based image enhancement*

Many methods work in the RGB color model. Histogram Equalization (HE) [47], Contrast Limited Adaptive Histogram Equalization (CLAHE) [48], Gamma Correction, and Generalized Unsharp Masking (GUM) [49] are regarded as the typical contrast enhancement methods to improve the global visibility of low-light images. Gray-World Assumption (GWA), White Balancing (WB) and Gray-Edge Assumption (GEA) are seen as traditional color correction methods to modify the color and saturation of the images. Due to the low energy of RGB components of underwater images (lacking of illumination in the underwater environments), it is frequent to introduce serious artifacts and halos, amplify the internal noise of the image and even cause color distortion when HE, GWA, WH and their variations are directly used for underwater image enhancement. Since the contrast of underwater images is low and the edge features are hazed, GEA often fails to enhance underwater images.

Fusion is an effective strategy of underwater image enhancement in single color model. In 2012, Ancuti et al. [50] proposed a fusion-based method. Firstly, two fusion images are generated from the input image: the first image is color corrected by white balance, and the second image is contrast enhanced by local adaptive histogram equalization. Then, four fusion weights are determined according to the contrast, salient features and exposure of the two fused images. Finally, the two fused images and the defined weights are combined to produce the enhanced images with better global contrast and detail information by using the multi-scale fusion strategy. In 2017, Ancuti et al. [32] introduced a new method for color balance and fusion for underwater image enhancement. Considering the underwater optical imaging theory, the proposed underwater white balancing aiming at compensating color cast caused by the light with selective attenuation is gamma corrected and sharpened to generate two fusion images and associated weight maps, which are merged based on the standard multi-scale fusion strategy. Their proposed enhanced images and videos are characterized by better exposedness of the dark regions, improved global contrast and edges sharpness.

In 2017, Liu et al. [51] proposed a method called Deep Sparse Non-negative Matrix Factorisation (DSNMF) to estimate the illumination of underwater images. First, the observed images were segmented into small blocks, each channel of the local block was reconstructed into a [R, G, B] matrix, and the depth of each input matrix was decomposed into multiple layers by the sparsity constraint of the DSNMF method. The last layer of the factorization matrix is used as the illumination for the patch, and the image is adjusted with sparse constraints. After factorization, the local block illumination of the original image is estimated to obtain the enhanced image.

*2) MCM-based image enhancement*

In 2005, Torres-Méndez et al. [52] used Markov Random Field (MRF) to describe the correlation between underwater images before and after distortion, and enhanced the color of images based on the maximum a posteriori. When calculating the dissimilarity of image patches, the image is transformed to CIE-Lab color space to represent equal perceived differences. The experimental data obtained from different underwater scenes verified the feasibility and effectiveness of this method.

In 2007, Iqbal et al. [53] proposed an underwater image enhancement algorithm based on an Integrated Colour Model (ICM). Firstly, the heavily attenuated GB channels in the RGB color model are stretched through the entire range [0, 255]. Then the image is converted to the HSI color model; and the *S* and *I* components are finally stretched with sliding histogram stretching to improve the saturation and brightness of the output image.

In 2010, Iqbal et al. [33] proposed an unsupervised Colour correction method based on Von Kries hypothesis (VKH) and contrast optimization of selective histogram stretching. UCM can effectively remove blue-greenish cast and improve brightness of low components. In 2015, Ghani et al. [34], [36], [54] adopted the Rayleigh distribution function to redistribute the input image in combination with the variation of ICM and UCM, improving image contrast and reducing over-enhancement, over-saturation region and noise introduction.

In 2017, Ghani et al. [37] put forward Recursive Adaptive Histogram Modification (RAHIM), which can increase the natural performance of image color by modifying saturation and brightness of the image in the HSV color model through Rayleigh distribution and the human visual system and finally the enhanced image is converted to RGB color model.

The Retinex theory simulates the mechanism of the human vision system as it perceives the world. The term of Retinex is created by the combination of the "retina" and "cortex". It attempts to achieve the color constancy when the scene is dominated by a certain illumination, which has a similar situation in the underwater environment. In 2014, Fu et al. [55] firstly proposed a simple RGB color cast correction algorithm for underwater images. Then, based on the theory of retina cortex, a new frame was proposed to separate direct light from reflected light in CIE-Lab color model. Finally, different strategies were used to highlight the separated light components to enhance the contrast of underwater images. In 2017, Zhang et al. [56] improve the above methods and extend the Retinex framework for underwater image enhancement. The brightness $L$ and color $a, b$ components are filtered by bilateral filter and trilateral filter to remove the luminance in Lab color model and suppress the halo artifacts.

In 2013, Hitam et al. [57] adjusted CLAHE and built the mixture contrast limited adaptive histogram equalization (Mix-CLAHE) to improve the visibility of underwater images. The CLAHE was applied to the RGB color model and the HSV color model to generate two images, which are combined by the Euclidean norm. Experimental results show that Mix-CLAHE can significantly improve the visual quality of underwater images by enhancing contrast, reducing noise and artifacts.

In 2018, Huang et al. [8] proposed relative global histogram stretching (RGHS) in RGB and CIE-Lab color models. The pre-processed image based on the theory of Gray-World employed adaptive histogram stretching in the RGB color model according to distribution characteristics of RGB channels and selective attenuation of light propagating under the water. Finally, the brightness $L$ and color $a, b$ components in the CIE-Lab color space are operated as linear and curve adaptive stretching optimization, respectively. RGHS can improve the visual effect of the image and retain available information by avoiding the blind enhancement on account of underwater image characteristics.

*B. TRANSFORM-DOMAIN IMAGE ENHANCEMNT*

In the frequency domain, the high-frequency image component usually corresponds to the edge region where the pixel values have great changes; whereas, the low-frequency component represents the flat background region in the image [58]. The transform-domain image enhancement methods commonly transform the spatial domain image into the frequency domain (e.g., through the Fourier Transform) [59], and improve the quality of underwater images by amplifying the high-frequency component and suppressing the low-frequency component, simultaneously [60]. The hazed underwater images often have the problem that the difference between the high-frequency component of the edge region and the low-frequency component of the background region is small [61]. Therefore, underwater image quality can also be improved by using transform-domain methods [62], such as homomorphic filter [63], high-boost filter, wavelet-transform, etc.

In 2010, Prabhakar et al. [64] used a homomorphic filter and an anisotropic filter to correct non-uniform illumination and smoothing the image. Finally, they applied adaptive wavelet sub-band thresholding with a modified BayesShrink function to implement de-noising.

Recently, underwater image enhancement methods based on Wavelet transformation have been used more often. In 2016, Amjad et al. [38] proposed a wavelet-based fusion method to enhance the hazy underwater images by addressing the low contrast and color alteration issues. Firstly, two fusion images are generated from the original image, by stretching the value component of the original image over the whole range in HSV color model and enhanced by CLAHE. Then, the wavelet-based fusion method consists of a sequence of low-pass and high-pass filters to eliminate unwanted low and high frequencies presented in the image, and acquire details of approximation coefficients separately for making the fusing process convenient.

In 2017, Vasamsetti et al. [40] proposed a framework of wavelet-based perspective enhancement technique for underwater images. Since changing the sign of a wavelet coefficient can result in undesirable modifications of an image, they applied the discrete wavelet transform (DWT) on the RGB channels to generate two decomposition levels, and collect the approximation and detailed responses for these parts to reconstruct the gray scale images for R-G-B channels. Meanwhile, this method can be used as the pre-processing of underwater detection and tracking techniques to boost the accuracy of the high-level underwater computer vision tasks.

Although the transform-domain underwater image enhancement methods can improve the visibility and contrast of the hazed images, they tend to over-amplify noise and cause color distortion.

*C. CNN-BASED IMAGE ENHANCEMNT*

In recent years, many studies have proved the effectiveness of deep learning methods in different application fields [65], such as image segmentation [66] and speech recognition [67]. Convolutional neural networks (CNN) are especially successful in image-based tasks – in fact several advanced deep learning models are based on CNN. There exist many results using diverse CNNs on low-level vision tasks [68] including image de-blurring [69]–[71], image de-raining [72], image de-noising [73], low-light image enhancement [74], [75] and image dehazing [76]–[79]. Yet very few methods are effective for underwater image enhancement [45].

In 2017, Perez et al. [80] proposed a CNN-based underwater image enhancement method, which trains an end-to-end transformation model between the hazed images and the corresponding clear images using pairs of degraded and recovered underwater images. Meanwhile, Wang et al. [44] also proposed an end-to-end, CNN-based underwater image enhancement framework, called UIE-net (Underwater Image Enhancement-net) for color correction and haze removal. The UIE-net adopts a pixel disrupting strategy to extract the inherent features of local patches of the image, which greatly fastens model convergence and improves accuracy. In 2018, Anwar et al. [81] used a database of synthetic underwater images that were produced in indoor environment to train a convolutional neural network (UWCNN), and used the UWCNN to reconstruct the clear underwater latent image directly. The generality of this model was verified with real and synthetic underwater images in a variety of underwater scenes.

Still, a large amount of training data is difficult to compile in deep sea environments, thus researchers used generative adversarial networks (GANs) [82] to generate realistic underwater images in an unsupervised pipeline. Li et al. [43] proposed WaterGAN to generate synthetic real-world images from in-air image and depth maps, then both raw underwater and true color in-air, as well as depth data were used to feed a two-stage deep learning network for correcting color-cast underwater images.

Similarly to waterGAN, Fabbri et al. [83] also adopted GANs to enhance underwater image. Firstly, they used CycleGAN to reconstruct distorted images based on the undistorted images, then the pairs of underwater images were fed to train a novel Underwater-GAN, which can transform hazed underwater images to clear and high-resolution images.

To relax the need for paired underwater images for network training and allow the use of unknown underwater images, Li et al. [84] proposed a weakly supervised underwater color correction model, which mainly consists of adversarial networks and multi-term loss function including adversarial loss, cycle consistency loss [85], and SSIM loss. This method can maintain the content and structure of the input underwater image but correct its color distortion. In 2019, Yu et al. [86] proposed Wasserstein GAN with gradient penalty term as the backbone network, designed the loss function as the sum of the loss of generative adversarial network and the perceptual loss and used a convolution patchGAN classifier as the discriminator of Underwater-GAN [83]. In 2019, Uplavikar et al. [87] proposed a domain-Adversarial learning-based underwater image enhancement, which can handle multiple types of underwater images and generate clear images by learning domain agnostic features.

So far, the reality of the generated underwater images has hardly been examined. To solve the difficulty in the development of CNN-based underwater image enhancement, in 2019, Li et al. [45] constructed a large-scale and real-world underwater image enhancement benchmark dataset (UIEBD), which was used to train a DUIENet that employs a gated fusion network architecture to learn three confidence maps.

## III. IFM-BASED IMAGE RESTORATION

Underwater image restoration usually establishes an effective degradation model by analyzing the underwater imaging mechanism and the basic physics of light propagation, then deduces the key parameters of the constructed physical model via some prior knowledge, and finally recovers the restored image by reserving compensation processing [88]. The simplified image formation model (IFM), given by equation (2) – Section I – is regarded as an effective and typical underwater image model for restoring underwater images. IFM-based restoration methods need to estimate two key optical parameters [89]: background light (BL) and transmission map (TM). In this section, we will introduce the prior-based and CNN-based image restorations, and explain how these recover natural colors of underwater images by estimating BLs and TMs.

### A. PRIOR-BASED IMAGE RESTORATION

Light absorption and scattering and suspended particles are the main causes of the underwater image degradation. With regards to the optical properties (e.g., selective light attenuation) or its representation (e.g., hazy effect), different prior-based methods were used or deducted for underwater image restoration. These include: dark channel prior (DCP) [13], [90]; underwater dark channel prior (UDCP) [91], [92]; maximum intensity prior (MIP) [93]; red channel prior (RCP) [94]; blurriness prior (BP); underwater light attenuation prior (ULAP) [95]; and others. According to these priors, the BL and TM (or depth map) can be derived and then be used into the IFM model for image restoration.

A summary of some mainstream prior-based underwater image restoration methods in the order of publishing year is given in Table 1. The table shows the BL estimation formula (Column 3), the TM estimation formula (Column 4), and their corresponding prior knowledge (Column 5, where the left and right sides of the slash represent the prior knowledge used in BL estimation and TM estimation, respectively). All parameters in Table 1 are simplified as follows.

$$P^c = min_{y \in \Omega(x), c \in \{r,g,b\}} \left(I^c(y)\right),$$
$$P^{c'} = min_{y \in \Omega(x), c' \in \{g,b\}} \left(I^{c'}(y)\right), \text{ and} \quad (3)$$
$$MIP^c = max_{y \in \Omega(x)}\left(I^r(y)\right) - max_{y \in \Omega(x), c' \in \{g,b\}}\left(I^{c'}(y)\right),$$

where $c \in \{r, g, b\}$, $c' \in \{g, b\}$, $B^c, t^c$ and $B^{c'}, t^{c'}$ represents BL and TMs of RGB channels and GB channels, respectively, $B^r, B^g, B^b$ and $t^r, t^g, t^b$ represent BL and TM of R-G-B channels. In the literature relating to MIP, $c'$ can be $\{g\}, \{b\},$ or $\{g, b\}$.

The following subsections describe the different types of priors used for underwater image restoration.

*1) DCP-based image restoration*
DCP, proposed by He et al. [13], is widely used for image dehazing. Due to the similarities between a hazed outdoor image and an underwater image, the DCP-based dehazing method is widely applied to underwater image enhancement.

The dark channel prior was based on the observation that clear day images contain some pixels which have very low intensities (close to zero) in at least one color channel. When directly using DCP for underwater image dehazing [96], the BL can be estimated in two steps: simply select the top 0.1% brightest pixels in the dark channel, and then among these pixels, select the pixels with the highest intensity in the input image. By minimizing both sides of the IFM model (Eq.(2)), the transmission map can be estimated.

**TABLE 1.** Formulas for estimation of BL and TM, and corresponding priors in underwater image restoration methods.

| Year | Methods | BL Estimation | TM Estimation | Prior |
|---|---|---|---|---|
| 2010 | [93] | $I^c(arg\ min\ MIP^c(x))$ | $t^c = MIP^c(x) + (1 - min\ MIP^c(x))$ | MIP/MIP |
| 2010 | [14] | $I^c(arg\ max\ P^c(x))$ | $t^c = 1 - min_{y \in \Omega(x),c}(I^c(y)B^c)$ | DCP/DCP |
| 2011 | [96] | $I^c\left(arg\ max_{x \in p_{0.1\%}} \sum_c I^c(x)\right)$ | $t^c = 1 - med_{y \in \Omega(x),c}\left(\frac{I^c(y)}{B^c}\right)$ | DCP/DCP |
| 2012 | [16] | $I^c(arg\ max\ P^c(x))$ | $t^c = 1 - min_{y \in \Omega(x),c}\left(\frac{I^c(y)}{B^c}\right), t^{c'} = (t^r)^{\frac{\beta^{c'}}{\beta^r}}$ | DCP/DCP |
| 2013 | [100] | $I^c\left(arg\ min\left(I_{dark}^r(x) - (max\ I_{dark}^{c'}(x))\right)\right)$ | $t^{c'} = 1 - min_{y \in \Omega(x),c'}\left(\frac{I^{c'}(y)}{B^{c'}}\right),$ $t^r = \tau max_{y \in \Omega(x)} I^r(y), \tau = \frac{avg(t^{c'})}{avg(max_{y \in \Omega(x)} I^r(y))}$ | MIP/UDCP |
| 2013 | [91] | $I^c(arg\ max\ P^c(x))$ | $t^c = 1 - min_{y \in \Omega(x),c'}\left(\frac{I^{c'}(y)}{B^{c'}}\right)$ | UDCP/UDCP |
| 2015 | [94] | $I^c\left(arg\ max_{x \in p_{10\%}} \sum_c I^c(x)\right)$ | $t^r = 1 - min\left(min_{y \in \Omega(x),c'}\left(\frac{I^{c'}(y)}{B^{c'}}\right), \lambda min_{y \in \Omega(x)} Sat(y),\right.$ $\left. min_{y \in \Omega(x)}\left(\frac{I^r(y)}{1-B^r}\right)\right), t^{c'} = (t^r)^{\lambda^{c^{g,b}}}$ | RCP/RCP |
| 2015 | [98] | $I^c(arg\ max_{x \in p_{0.1\%},c'}|I^r(x) - I^{c'}(y)|)$ | $t^r = 1 - min_{y \in \Omega(x),c}\left(\frac{I^c(y)}{B^c}\right), t^{c'} = (t^r)^{\frac{\beta^{c'}}{\beta^r}}$ $\frac{\beta^{c'}}{\beta^r} = \frac{B^{r,\infty}(m\lambda^c + i)}{B^{c',\infty}(m\lambda^r + i)}$ | DCP+MIP/DCP |
| 2015 | [99] | $\frac{1}{|p_{0.1\%}|} \Sigma_{x \in p_{0.1\%}} I^c(x)$ | $t^c = F_g\{C_r[P_r(x)]\}$ | DCP/BP |
| 2016 | [102] | $Avg\left(I^{c'}(arg\ min MIP^c)\right)$ | $t^c = 1 - min_{y \in \Omega(x),c'}\left(\frac{I^{c'}(y)}{B^{c'}}\right)$ | MIP/UDCP |
| 2017 | [105] | $\alpha B_{max}^k + (1-\alpha) B_{min}^k$ | $\theta_b[\theta_a d_D + (1-\theta_a)d_R] + (1-\theta_b)d_B$ | IBLA/IBLA |
| 2018 | [95] | $I^c(arg\ max_{x \in p_{0.1\%}} d(x))$ | $t^c = Nrer_c^{d(x)}, d(x) = ULAP(x)$ | ULAP/ULAP |

In 2010, Chao et al. [14] directly used the DCP to recover the underwater images and remove the scattering of underwater images, respectively. But the restored images show a limited improvement and even suffer from additional color distortion compared with the original images. In 2011, Yang et al. [96] proposed a DCP-based fast underwater image restoration method to reduce the complexity of computation execution. They mainly replaced soft matting with median filtering to estimate the depth information of images and finally used a color correction to improve the contrast and brightness of restored images. This is only suitable for the underwater images that have rich colors, and cannot recover underwater images with color cast or dim scene.

Some studies aimed to refine the DCP-based parameter estimations. In 2012, Chiang et al. [16] proposed wavelength compensation and image dehazing (WCID) to remove the artificial light, compensate three channels with different attenuation characteristics, and eliminate the effect of the haze combined with the classical DCP algorithm. In 2014, Serikawa et al. [97] combined the DCP with fast joint trigonometric filtering (JTF) to compensate the attenuation discrepancy along the propagation path. The JTF can improve the TM estimated by the traditional DCP to ease the scattering and color cast, reduce the noise level of the image, and improve image contrast and edge information. In 2015, Zhao et al. [98] derived IFM-based underwater inherent optical properties from the background color of underwater images. They revealed the attenuation coefficients of RGB channels based on the relationship between BL and inherent optical properties. The traditional DCP was used to estimate the TM of R channel, and then the TMs of GB channels were derived by considering the exponential relationship with the attenuation coefficient. In 2015, Peng et al. [99] picked up the top 0.1% brightest pixels in dark channel and then selected the average value of the corresponding intensities in the input image as final background light.

The dark channel prior is easily affected by the selective light attenuation in underwater environments, thus many underwater-specific DCP were developed.

*2) Underwater DCP-based image restoration*
As the red light attenuates much faster than the green and blue lights when it propagates in water, the red channel of an underwater image will dominate in the dark channel. To eliminate the influence of red, underwater dark channel prior (UDCP) [91], proposed by Drews et al. in 2013, only considers GB channels to produce underwater DCP. In the meantime, Wen et al. [100] proposed a new underwater optical model similar to the UDCP and used it to estimate the scattering rate. Although the proposed UDCP can obtain more accurate TM than DCP, the restored images are still not satisfactory because these methods ignore the imaging characteristics of the R and GB channels, and may not work well in turbid water.

In 2015, Lu et al. [101] found that the lowest pixel value of the RGB channels in turbid water is not always the red channel, but is occasionally the blue channel, and the blue channel is absorbed the least. Hence, they used a dual dark channel (red and blue) to estimate coarse TM and reduced the halo and mosaic effects of the course TM by a weighted median filter. In 2016, Li et al. [102] proposed single underwater image restoration by UDCP-based GB channels dehazing and R channel color correction based on the Gray-World hypothesis, and then took adaptive exposure map to balance the overall color of the restored images. In 2015, Galdran et al. [94] proposed an automatic red channel underwater image restoration based on red channel prior (RCP), which exacts the dark channel from reversed red channel and blue-green channels. Meanwhile saturation information of hazed images was introduced to adjust TM to effectively enhance the artificial light region and improve the overall color fidelity of images. However, the colors of some restored images present visually incorrect and unreal.

*3) MIP-based image restoration*

By discovering the strong difference in attenuation between the R and the GB channels of underwater images, Carlevaris et al. [93] proposed a novel prior knowledge for scene depth estimation, so-called maximum intensity prior (MIP). The MIP method defined TM by the difference between the maximum R channel intensity and the maximum of the G and B channels, and a shift from the closest point in the foreground represented by the largest difference between color channels. Experimental results showed that MIP could describe coarse depth maps of images.

The MIP was also used for BL estimation. In 2013, Wen et al. [100] adopted and modified the MIP to estimate the BL of underwater images, with the assumption that the intensity of the R channel was relatively lower than that of the GB channels in the background area. In 2015, Zhao et al. [98] estimated BL based on DCP and MIP. They firstly picked up the top 0.1% brightest pixels in the dark channel, and then selected the pixel with the maximum difference of B-G channels or G-R channels among these pixels. In 2016, Li et al. [102] determined the background light directly selected from the pixels of the maximum difference. With a mixing strategy in [103], [104], they firstly selected one flat background region based on the quad-tree subdivision, and then picked up the top 0.1% brightest pixels in the dark channel from the candidate region, and finally chose one of these pixels with the maximum difference of R- B channels in the original image as the global background light.

*4) Other prior-based image restoration methods*

Besides the priors mentioned above, there are some priors that are not widely used but work effectively for underwater image restoration.

In 2015, Peng et al. [99] proposed blurriness prior (BP) based on the assumption that the deeper the scene depth was, the more blurred the underwater object, and then used BP to estimate scene depth and completed image restoration. In 2017, Peng et al. [105] further improved the BP and proposed image blurring and light absorption (IBLA) to estimate more accurate background light and underwater scene depth, and restored underwater images under various types of complicated scenes.

In 2018, Peng et al. [106] proposed a generalized dark channel prior (GDCP) based on the depth-dependent color by calculating the difference between the observed intensity and the background light, which can be used to estimate ambient light and scene transmission map. To reduce information loss when recovering the natural underwater images, in 2016, Li et al. [103] represented the TM of the most degraded R channel based on the minimum information loss principle (MILP). Then, the histogram distribution prior, that is the average histogram distributions of natural-scene images, was used as the template to adjust the contrast and brightness of underwater images. In 2017, considering the attenuation effect of absorption and scattering strongly correlated with imaging depth, Wang et al. [107] proposed maximum attenuation identification (MAI) to derive the background light and depth map from degraded underwater images.

In 2018, Song et al. [95] proposed a rapid and effective scene depth estimation model based on underwater light attenuation prior (ULAP), which assumed the difference between the maximum value of G-B intensity and the value of R intensity in one pixel of the underwater image strongly related to the change of the scene depth. Based on the ULAP, a linear model was established to rapidly obtain scene depth map, which can be used to estimate the background light (BL) and transmission maps (TMs) for R-G-B channels are easily estimated to recover the true scene radiance under the water. which use instance data to extract some valuable feature vectors.

In 2017, Ding et al. [108] first used a new white balance algorithm to improve the overall quality of original underwater images, then adopted traditional CNN to estimate BL and TM of underwater images after color correction, and finally restored the underwater images based on IFM. Because pre-processed underwater images through color correction

*B. CNN-BASED IMAGE RESTORATION*
IFM-based underwater image restoration methods estimate the BLs and design of TMs by prior knowledge, condition assumptions and theory. With the rapid development of deep learning in image restoration, researches have already seen a significant change from parameter selection completely by artificial optimization models to automatic training models, lose the imaging characteristics of underwater environments, there will be over-saturated and over-enhanced areas in the restored image. In 2018, referencing to the multi-scale deep network put forward by Eigen [109], Cao et al. [110] stacked a coarse global CNN network and a refined network to estimate the background light and predict scene depth map. This method claimed a good recovered result, better than the existing image restoration method based on the IFM.

In 2018, Barbosa et al. [111] considered that the existing end-to-end framework may fail to enhance the visual quality of underwater images in lack of ground truth of the scene radiance. Hence they proposed a CNN-based method by using a set of image quality metrics to guide the restoration learning process without requiring ground truth data. Experiments showed that Barbosa et al's method improved the visual quality of underwater images, preserving their edges. In Hou et al.'s work [112], the prior knowledge and data information were aggregated to investigate the underlying underwater image distribution to correct color, by means of a data-driven residual architecture for transmission estimation and a knowledge-driven scene residual formulation for underwater illumination balance.

CNN-based image restorations estimate BLs or depth maps through feature learning. The performance of the methods relies on both network architecture design and training data. Due to the use of synthetic underwater images and potential defect of deep-learning architecture, these trained network models may only adapt to some limited types of underwater images. Compared with physical model and non-physical model, the deep learning method is time-consuming under the same restoration environment.

## IV. QUALITY IMPROVEMENT METHODS FOR UNDERWATER IMAGES: EXPERIMENTAL COMPARISONS

To study current development of quality improvement methods for underwater images, we firstly introduce image quality assessment metrics, and then take comprehensive comparisons on mainstream IFM-based underwater image restoration methods and IFM-free underwater image enhancement methods from both subjective and objective perspectives. Since BL and TM estimation determine the robustness and effectiveness of IFM-based methods, we also evaluate prior-based BL estimation models and prior-based TM estimation models, discussing the advantages and disadvantages of BL and TM estimation models, and the effect of BL results on TM estimation.

### A. THE METHODS TO BE COMPARISED

The compared methods of IFM-free image enhancement include: HE [47], CLAHE [48], integrated colour model (ICM) [53], unsupervised colour correction method (UCM) [33], Fusion-based underwater image enhancement method (Fusion-based, FB) [50], underwater image enhancement method based on Rayleigh distribution (RD) [113], and relative global histogram stretching (RGHS) [8].

The compared IFM-based underwater image restoration methods are: single image removal (SIR) based on dark channel prior (DCP) [13], initial underwater image dehazing (IUID) based on maximum intensity prior (MIP) [93], DCP-based rapid image restoration (RIR) [96], Underwater Transmission Estimation of Underwater Image (TEoUI) [91], underwater image restoration based on the new optical model (NOM) [100], underwater image restoration based red channel prior (RCP) [94], image blurriness and light absorption (IBLA) [105], and underwater light attenuation prior (ULAP) [95].

In order to ensure the fairness of each evaluation system, all test underwater images are pre-processed at the size of 400×600 pixels and processed by the compared methods with default parameters. All methods are implemented on a Windows 7 PC with Intel(R) Core(TM) i7-4790U CPU@3.60GHz, 8.00GB 1600MHz DDR3 Memory, running on Python3.6.3.

### B. IMAGE EVALUATION METRICS

Image quality can usually be affected by the optical performance of imaging equipment, instrument noise, imaging conditions, image processing and other factors. Image quality assessment (IQA) is often divided into subjective qualitative assessment (SQA) and objective quantitative assessment (OQA).

SQA is mainly dependent on the human visual system (HVS) to gain subjective impression of images. A proper SQA needs repeating a number of experiments (varying the factors that affect image quality) to generate a dataset, which is then scored by human observers, striving for statistical significance. Due to the low efficiency and complicated operation of SQA, in this paper we simply present the representative results from different image enhancement/restoration methods as the basis of subjective analysis.

OQA establishes mathematical model based on the HSV to calculate a quality index. Provided that accurate models are utilized, this method is significantly more efficient than SQA, since a vaster dataset may be scrutinized automatically. OQA is often divided into three kinds of image quality evaluation indexes: the full-reference (FR), the reduced-reference (RR) and the non-reference (NR) methods. When evaluating the quality of an image, FR and RR image quality metrics require or partially require a high-quality reference image. Unfortunately, the dehazed and natural reference image cannot be obtained in complicated underwater environment, unless there are synthetic images or when color boards in the terrestrial scene are taken into the underwater scene. In addition, due to the complicated underwater environment and optical imaging mechanism, the evaluation metrics for underwater images are limited. To fully understand the performance of the compared underwater image quality improvement methods, we choose multiple NR metrics developed for both specific underwater images and for general images, considering the aspects of information richness, naturalness, sharpness, and the overall index of contrast, chroma and saturation.

Entropy is interpreted as the average uncertainty of information. When applied to images, entropy represents the abundance of information observed from the image. When the contrast of the image is more uniform, the entropy is relatively higher, the better the quality of image will be and the clearer the image will be, otherwise the image with low contrast, whose pixel values are distributed within a small range, has smaller entropy and appears hazed.

Natural image quality evaluator (NIQE) [114] was established according to human vision sensitivity to high-contrast areas in images. It uses multivariate gaussian (MVG) to establish the feature model of sensitive areas, where the larger the values of these parameters are, the higher the image quality will be. A smaller score of NIQE indicates better perceptual quality.

Blind/Referenceless Image Spatial Quality Evaluator (BRISQUE) [115] measures image naturalness based on measured deviations from a natural image model, which is based on natural scene statistics. BRISQUE can represent the possible loss of image naturalness caused by distortion, whose range is from 0 to 100, and the bigger value is, the worse the image quality is.

In 2015, Yang et al. [116] discovered the correlation between sharpness and color of image and the subjective image quality perception and proposed an image quality evaluation method specially for underwater images, the underwater color image quality evaluation (UCIQE). UCIQE is a linear model of contrast, chroma and saturation in CIE-Lab color space, can be expressed as:

$$UCIQE = c_1 \times \sigma_c + c_2 \times con_l + c_3 \times \mu_s \quad (4)$$

where $\sigma_c$, $con_l$ and $\mu_s$ represent the standard deviation of image chromaticity, contrast of image brightness and average of image saturation, $c_1, c_2, c_3$ represent the weights of these parameters. Similar as UCIQE, underwater image quality measure (UIQM) [117] constructed the linear combination of underwater image colorfulness measure (UICM), underwater image sharpness measure (UISM) and underwater image contrast measure (UIConM). Thus, the larger the UCIQE and UIQM are, the better the underwater color image quality will be.

## C. ASSESSMENT ON OPTICAL PARAMETERS of IFM-BASED METHODS: BL & TM

*1) Comparisons of BL estimation models*

The BL estimation method for underwater images is often ignored by researchers. It determines the color tone and visual effect of restored images. Many estimation algorithms of TM also depend on the result of BL estimation to a large extent, which can be seen from Table 1. Thus, it is essential for IFM-based underwater image restoration to carry out a comparison of different BL estimation models. This section evaluates the performance of different BL estimation methods through subjective and objective performance analysis.

In order to compare BL estimation methods based on different priors, this review selects four typical images including shallow-sea fishes under natural light source, cliff under low-brightness scene and wrecked ships, and the swimming batfish in the foreground area, as exemplified in Fig 3 (a). The ground truth BLs of these test images in Fig 3(b), were produced from 15 people's manually annotation based on the principle of selecting the farthest point from the camera and the light used to illuminate the background area, as detailed in our previous work [25].

Fig 3 (c-m) show the BLs estimated by the methods with different priors. Among them, Fig 3 (c-e) show estimated results of DCP-based methods, Fig 3 (f) based on DCP and MIP, and Fig 3 (g-i) based on MIP only. Fig 3 (j) and Fig 3 (k) are UDCP-based and RCP-based BL estimation results, respectively. Fig 3 (l) and Fig 3 (m) are Fusion-based and ULAP-based BL results respectively.

By comparing Fig. 3 (c-e) with Fig. 3 (b), it can be seen that DCP-based BL estimation methods that partially choose the brightest pixel values in the whole image as final BLs, are often wrong. Some studies [13], [14], [96] have shown that DCP-based BL estimation methods can avoid blindly selecting the strongest pixel as the final BL, but DCP ignores the optical imaging characteristics of underwater images where significant difference between the R channel and the GB channels exists, leading to the failure of the DCP-based BL estimation method. Based on the maximum difference of R channel and GB channels in the background area, MIP can effectively avoid the interference of natural light source and over-bright foreground. Therefore, most results in Fig. 3 (g-i) are close to the ground truth BLs. But the method using MIP upon DCP, whose results are in Fig. 3 (f), generates much brighter BLs than the ground truth. This makes MIP useless and fails the estimation eventually.

The results of UDC in Fig. 3(j) are similar to those of the DCP-based BL estimation methods, because UDCP still ignores the great attenuation in R channel. Since RCP considers the dark region in the R channel as the BL candidate region, the BLs estimation is correct except for the cliff image in Fig. 3 (k). This is because the cliff image has very dark regions where the R component is very low.

The fusion-based BL estimation model selects three common estimated BLs as candidate BLs and determines the final BL based on selective weighted fusion. The results in Fig. 3(l) are better than DCP and UDCP-based methods. ULAP considers that the difference of R channel and GB channels is strongly correlated with the scene depth, and chose the BLs from the values of the farthest point of the original image. The output shown in Fig.3 (m) are close to the ground truth.

To quantitatively assess the results of BL estimations based on different prior knowledge DCP, MIP, UDCP, RCP, IBLA and ULAP, we computed BLs of 300 underwater images with these methods, and calculated the absolute differences between the estimated and the ground truth BLs with the tolerance of R channel and GB channels set as 30 and 40, respectively. That is to say, as long as the absolute difference is within the tolerance, the estimation is considered as correct. The BL accuracy is represented by the accumulated correct ratio of all test image BLs. Fig 4 shows comparisons of accuracy of BL estimation methods based on the different priors.

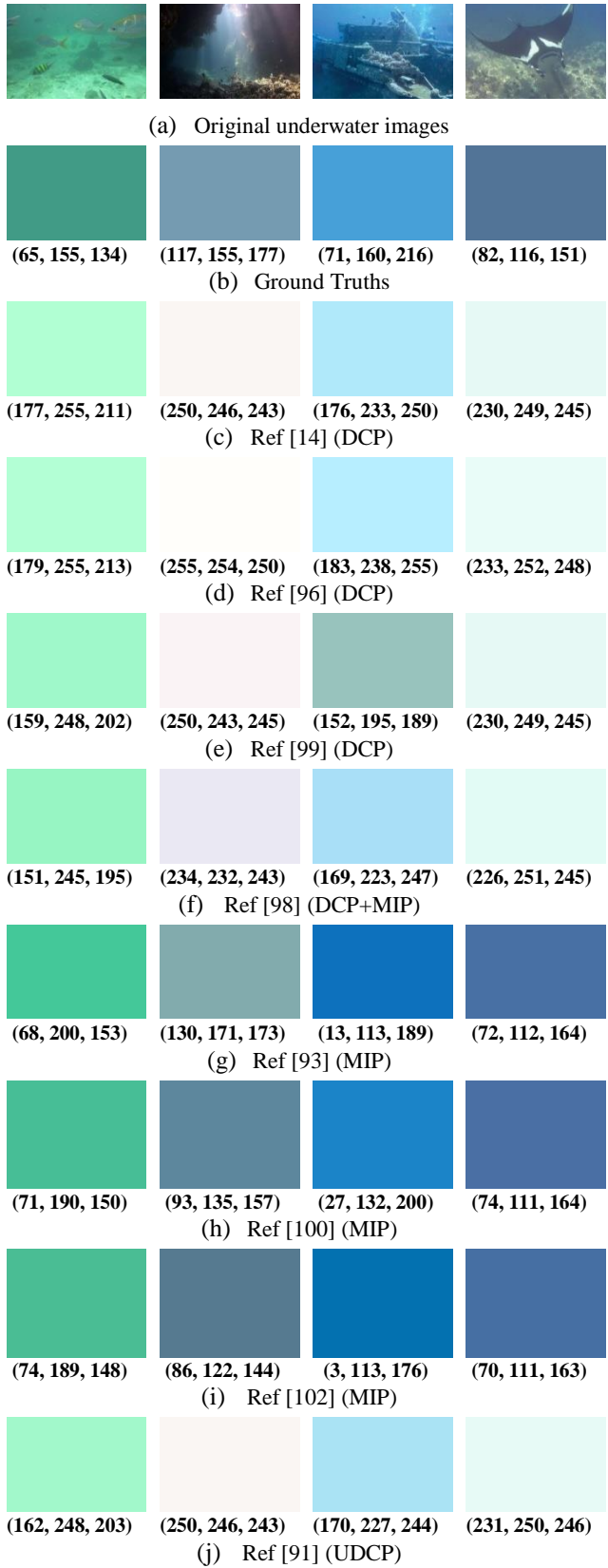

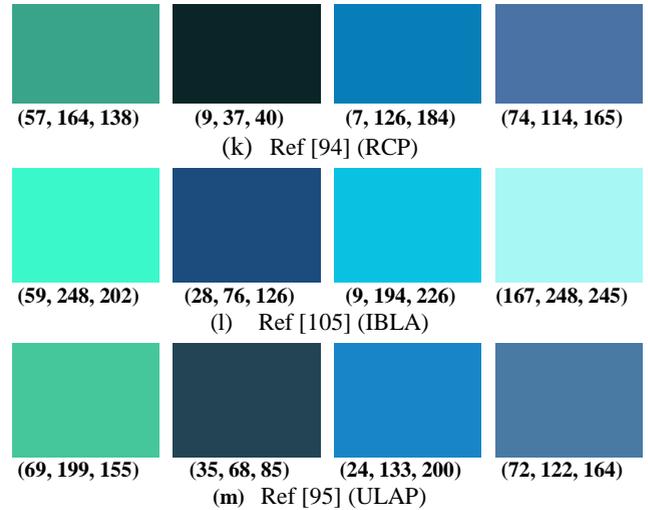

**FIGURE 3.** Comparisons of BLs estimation methods based on different priors.

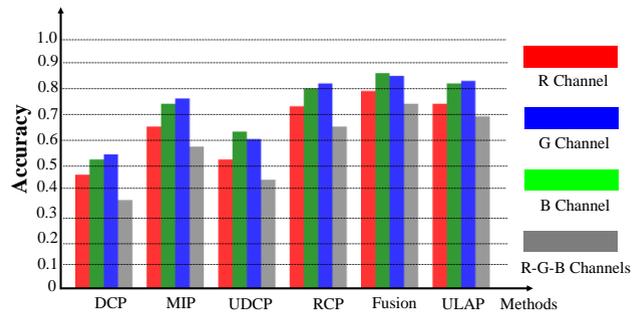

**FIGURE 4.** Comparisons of accuracy of BL estimation.

According to Fig. 4, the accuracies of DCP-based and UDCP-based BLs estimation are the lowest, indicating that DCP and UDCP are not suitable for estimating various types of underwater images. This conclusion is consistent with those presented in Fig. 3. Although MIP can successfully estimate BLs of the images in Fig. 4, its overall BL estimation accuracy on various types of underwater images is relatively lower. RCP-based, Fusion-based and ULAP-based BL estimation methods show better estimation results in the test dataset, but their accuracy across three R, G and B channels is still lower than 80%.

### 2) Comparisons of TM estimation models

When comparing the accuracy of transmission map (TM) estimation models, this review analyzes the correctness of TM estimation through subjective assessment due to non-reference depth/transmission map. The closer an object is to the camera, the higher its TM valuate is, and the whiter it shows on the TM or depth map. On the contrary, the farther, the darker. This principle is used to evaluate the performance of TM estimation methods based on different priors.

Four challenging underwater images are selected from the underwater image dataset as test images. As shown in Fig. 5 (a) from left to right, the first image of cliff and the third of coral have complex scenes but different color tones; the second of a shoal is hazy and with many fishes to distinguish; and the fourth has rocks with artificial light spot. Fig. 5 (b-

VOLUME XX, 2019    11

m) demonstrate the estimated R channel results based on the TM estimation models with different priors. In order to better represent the results of TM estimation models, this review refines all the estimated TMs by guided filter [118].

In Fig. 5 (b-e), DCP-based TM estimations perform poorly, except a relatively reasonable result for the shoal image. They are hardly distinct the depth of objects in a complex scene (e.g., the cliff and coral images) and mistake the light spot as the farthest when artificial light exists. The problem is mainly due to the false BLs estimation. The DCP-based method will choose the whitest point as the far background area, which can be either a white object or light spot. With this BL as the referenced far area, the complete depth map will be wrong. In general, DCP-based TM estimation methods are sensitive to underwater images with different characteristics and have a low applicability.

As can be seen from Fig. 5 (f), MIP proposed for underwater imaging characteristics can roughly estimate the TMs of the four images, but the estimated values are large. This leads to an overall white map and fuzzy image details.

Compared with Fig. 5 (h-i), Fig. 5 (g) shows the incorrect TMs of the first three images and relatively correct TM of the last image because the TM of R channel is estimated based on the local maximum values of R channel, TMs of G-B channels are estimated based on the UDCP. However, the TMs of R channel is estimated based on UDCP in Fig. 5 (h-i). In Fig. 4 (j), RCP-based TM estimation method effectively uses saturation information to avoid the influence of artificial light on the TM estimation, but the estimated TMs is too large to be used to restore the image. In Fig. 5 (j), the IBLA-based TM estimation methods are applicable to the underwater images with four typical features, and highlight the local details and texture information of TMs of the underwater image. In Fig. 5 (l), the ULAP-based TM estimation method underestimates TM of shoal, but this linear model of TM estimation can quickly estimate the TMs of the remaining three images, especially the area where the artificial light source exists.

To sum up, DCP-based TM estimation methods are applicable to some underwater images, but are easy to cause TMs of G-B channels much larger than actual values. MIP-based TM estimation methods can coarsely estimate the depth information of the original images, but need to refine the details of TMs. The UDCP-based TM estimation method can directly avoid the influence of the R channel on TM estimation, and improves the DCP-based TM estimation method to a certain extent. The RCP-based TM estimation model can obtain the approximate depth information, but the overall estimation of TM is too large to work. The IBLA-based TM estimation method, by considering both the light attenuation and image blurriness of underwater images, gains more accurate TM estimations, but its computing complexity is high. The ULAP-based TM estimation is influenced by the objects whose intensity difference between R channel and GB channels are significant high, but is not affected by artificial light source.

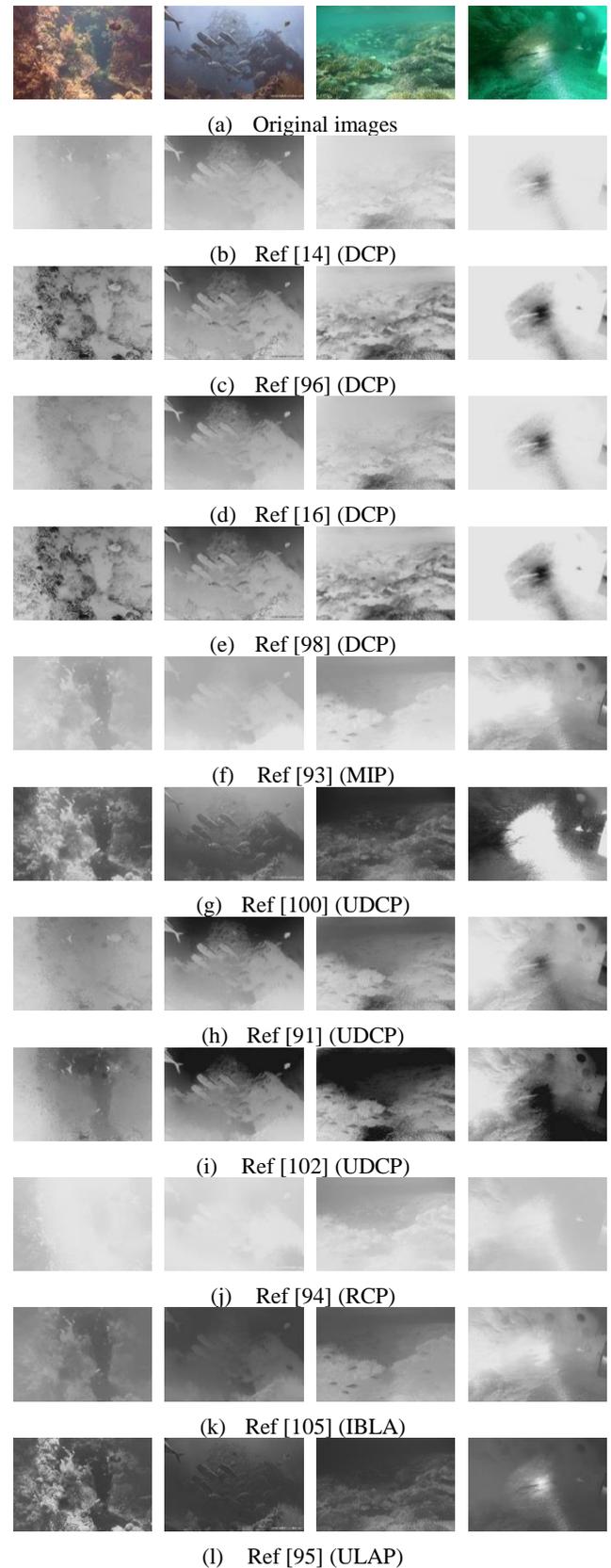

(a) Original images

(b) Ref [14] (DCP)

(c) Ref [96] (DCP)

(d) Ref [16] (DCP)

(e) Ref [98] (DCP)

(f) Ref [93] (MIP)

(g) Ref [100] (UDCP)

(h) Ref [91] (UDCP)

(i) Ref [102] (UDCP)

(j) Ref [94] (RCP)

(k) Ref [105] (IBLA)

(l) Ref [95] (ULAP)

**FIGURE 5.** Comparisons of accuracy of TM estimation.

## D. OVERALL PERFORMANCE OF UNDERWATER IMAGE ENHANCEMENT AND IMAGE RESTORATION: EVALUATION AND DISCUSSION

In this section, we evaluate the overall performance of IFM-free and IFM-based underwater image improvement methods, given in Section IV.A. As a benchmark, we have adopted a dataset including four types of underwater images, which is commonly used in the literature. This include one relatively clear scene and three challenging underwater images under the greenish, turbid and low-visibility scene (Fig. 6(a)). Subjective and objective analysis are employed on the enhanced images. For the IFM-based methods, the estimated BLs and TMs are also demonstrated to aid the discussion.

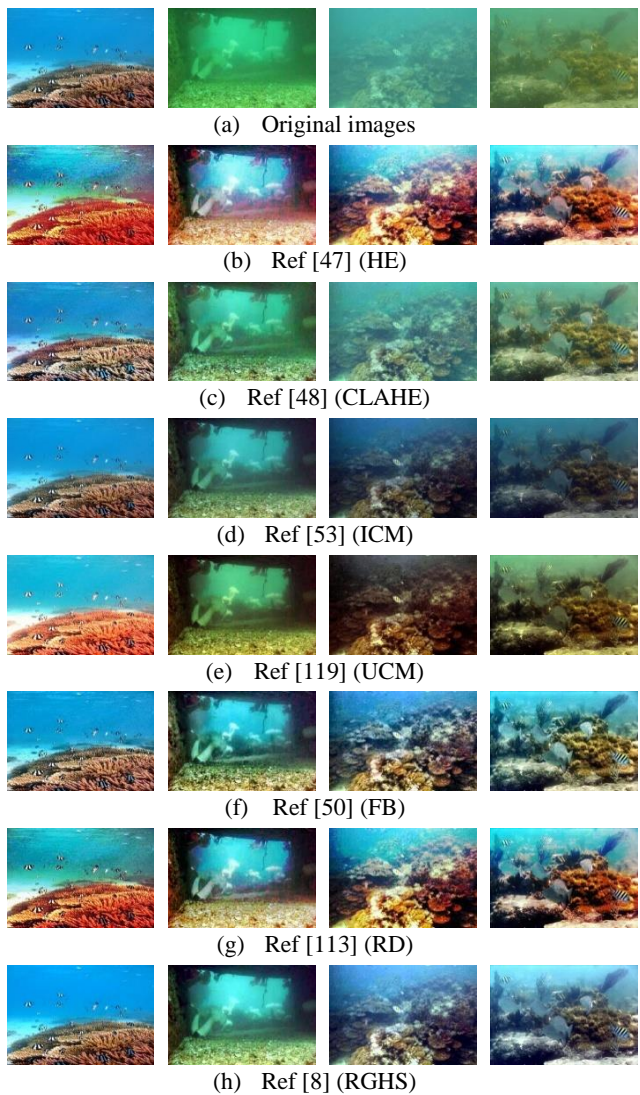

**FIGURE 6** Comparisons on results of IFM-free image enhancement methods.

*1) Subjective Analysis*
Fig. 6 (b-h) shows the results of IFM-free image enhancement methods. The images enhanced by HE method (Fig. 6 (b)) present an overwhelming red tone and amplify the noises of the original image. Both CLAHE and RGHS are based on adaptive parameters to avoid a global histogram stretching or blind pixel redistribution to reduce sharpness. Thus their results in Fig. 6 (c) and Fig. 6 (h) are not over-enhanced.

RGHS shows a better dehazing effect than CLAHE. ICM and UCM redistribute the S and I components in HSI color space, may lead to under- and over-saturated images, as shown in Fig. 6 (d-e). RD modified ICM and UCM by combining with Rayleigh distribution in the HSV color model to minimize under- and over-enhanced areas of output images. But RD veils local detailed information of the enhanced images. Although Fusion-based (FB) image enhancement method can significantly improve contrast and chromaticity of images, while the noise is also inevitably introduced to the enhanced images. Overall, the IFM-free methods can effectively improve contrast, visibility and luminance of the underwater image, but bring unnatural chroma and enlarged noises.

Fig. 7 shows the estimated BLs, TMs, and restored results of IFM-based image restoration methods. SIR directly applying DCP to BL and TM estimation of underwater images, has failed to estimate the parameters and results in the failure of restoration, as shown in Fig. 7 (a). But this does not affect the clear underwater image because the estimated TM falls flat.

Yet, in Fig. 7(b), RIR generates distorted images, especially appearing with a reddish tone, because it considers that TMs of R-G-B channel are the same. In Fig. 7(c), IUID based on MIP obtains correct BLs but cannot remove haze nor correct color cast for the three challenging images. This can be explained by its estimated TMs that cannot distinguish the scene depth.

TEoUI based on UDCP obtains relatively reasonable TMs, but the values of BLs are too big to make the clear underwater image oversaturated. When producing TMs, NOM uses the median filter to replace soft matting to improve the operation efficiency. However, according to Fig. 7 (e), it over-enhances the R component, under-enhances the G-B components, and introduces a large amount of noise, and finally causes distortion.

The best restoration images are produced by IBLA and ULAP, as shown in Fig. 7 (f-g). IBLA and ULAP can adopt characteristics of underwater light attenuation to obtain the correct depth map or TMs of R-G-B channels by building the optical imaging relationship of R-G-B channels.

To sum up, the current IFM-based underwater image restoration methods can only complete basic dehazing work but cannot effectively deal with color restoration for various underwater images, but the color correction can be imported as post-processing to improve the brightness, color and contrast of restored images. According to results of different restoration methods, our review raises one question whether this simplified imaging formation model is actually inappropriate for underwater image restoration.

*2) Objective Analysis*
Underwater image restoration/enhancement is to improve the visibility, color and saturation of images, and reveal

VOLUME XX, 2019　　　　　　　　　　　　　　　　　　　　　　　　　　　　　　　　　　　　　　　　　　　　13

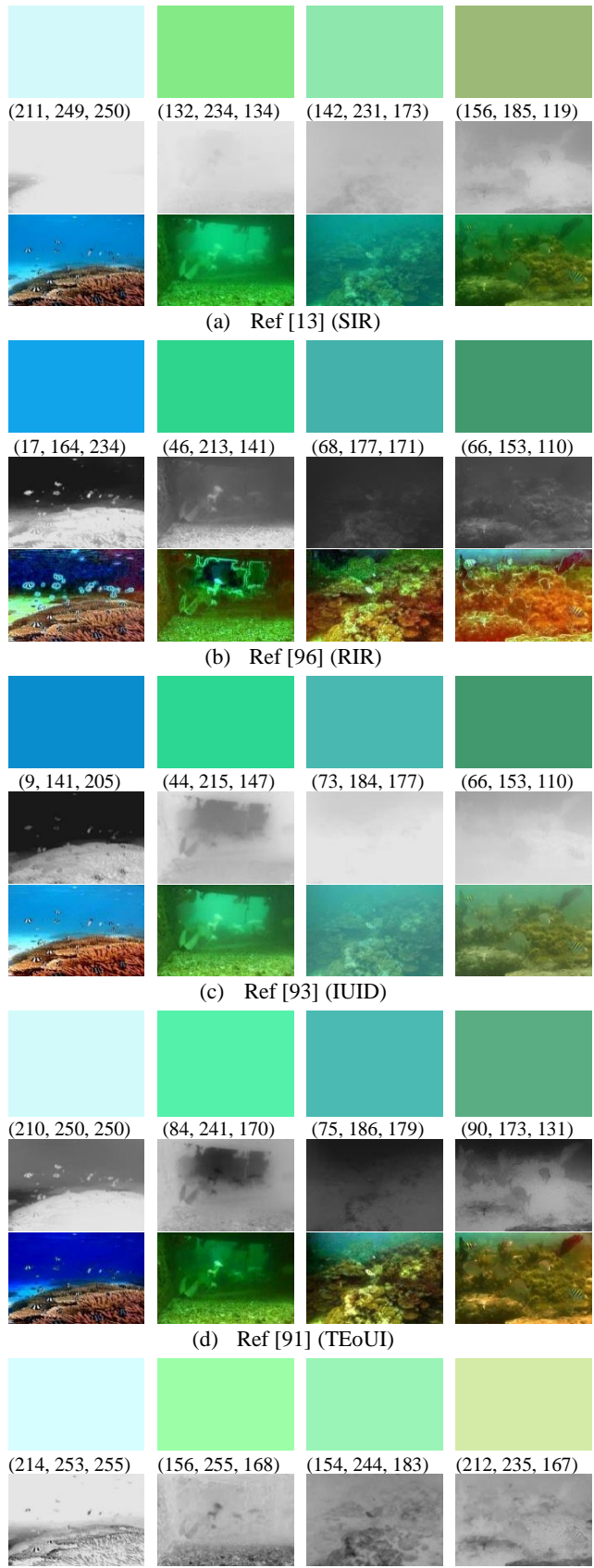

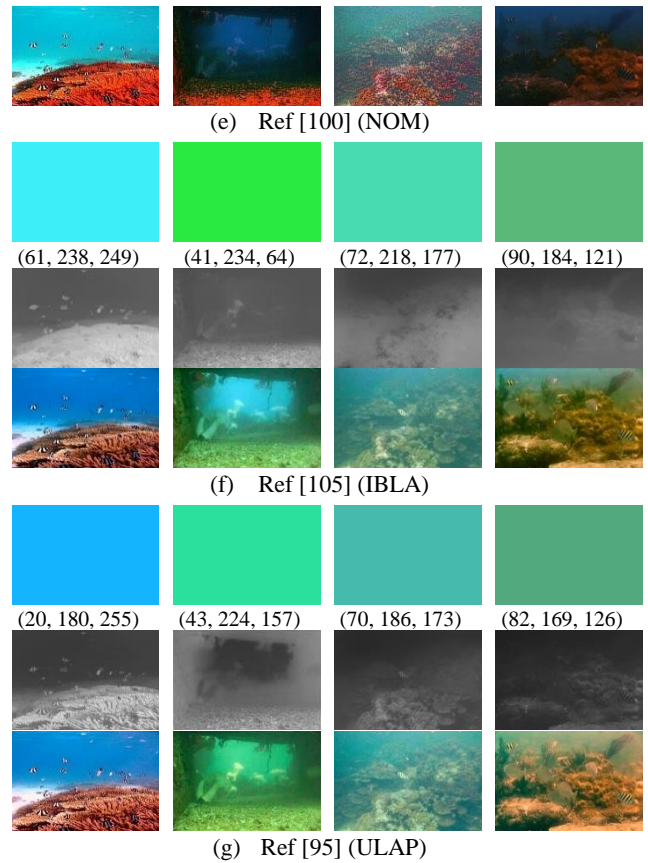

**FIGURE 7.** Comparisons on BLs, TMs and restored images of IFM-based image restoration methods.

detailed information, for the purpose of feature extraction and computer vision analysis. Due to the absence of reference underwater images (ground truth), this review selects five kinds of non-reference image quality metrics to quantify information entropy, distortion and the balance of brightness, contrast and color for underwater images. The five metrics are ENTROPY, BRISQUE, NIQE, UIQM and UCIQE, as introduced in Section IV.B above.

Table 2, shows the average values of the five quantitative evaluations of the restored and enhanced images, highlighting the best results in bold. Entropy values of the IFM-free results are generally higher than those of the restored images by IFM-based methods. This suggests that image enhancement algorithms can improve the information abundance contained in the image. Yet, image enhancement algorithms blindly amplify the useless information, especially the noises, as per Fig 6 (b-h). Although the entropy values of the enhanced images obtained by HE is the highest, it can be seen that enhanced images appear unnatural according to Fig 6 (b).

Both BRISQUE and NIQE models are built using outdoor images as evaluation criteria. TEoUI results in Fig 7 (d) are obviously unnatural, perceptually, but obtain the best naturalness (the lowest BRISQUE score). By contrast, FB, IBLA and ULAP obtain relatively higher BRISQUE score. SIR gains the best assessment according to NIQE, but it gives almost no improvement to the original underwater images. Therefore, it is problematic to directly use the quality assessment metrics based on outdoor images to evaluate underwater images.

UCIQE and UIQM were developed to reflect the quality of underwater color images. According to these two metrics, overall IFM-free methods perform significantly better than IFM-based methods. But these underwater image quality assessment metrics favor the images with high contrast and extreme chroma, such as the images produced by HE and NOM (shown in Fig. 6(b) and Fig. 7(e), respectively). Both UCIQE and UIQM metrics focus on the intensities of low-level features such as contrast, chroma and saturation but ignore higher semantic or prior knowledge from human perception.

**TABLE 2.** Quantitative analysis of restored and Enhanced results based on different methods.

| Compared methods | | Image Quality Assessment Metrics | | | | |
|---|---|---|---|---|---|---|
| | | ENTROPY | BRISQUE | NIQE | UIQM | UCIQE |
| IFM-free underwater image enhancement | HE | **7.8139** | 28.6079 | 3.9654 | 4.0399 | 0.6818 |
| | CLAHE | 7.1132 | 27.3445 | 3.6338 | 2.0644 | 0.6567 |
| | ICM | 6.9117 | 33.1758 | 3.4253 | 2.2999 | 0.5872 |
| | UCM | 7.2643 | 28.2424 | 3.6339 | 3.3228 | 0.6131 |
| | FB | 7.5269 | 32.9730 | 3.9176 | 2.7567 | 0.6684 |
| | RD | 7.7487 | 29.0286 | 3.7631 | 3.2654 | 0.6721 |
| | RGHS | 7.4759 | 28.3178 | 3.5161 | 2.0116 | 0.6176 |
| Avg(Var) | | 7.04(0.09) | 29.67(4.86) | 3.69(0.03) | 2.82(0.49) | 0.64(0.001) |
| IFM-based underwater image restoration | SIR | 6.3973 | 33.5067 | **3.3175** | 0.1605 | 0.5054 |
| | IUID | 6.5484 | 29.6948 | 3.3645 | 0.7895 | 0.5270 |
| | RIR | 6.4863 | 27.5484 | 4.2616 | 2.5178 | 0.5578 |
| | NOM | 7.3464 | 33.2872 | 4.3518 | **4.1640** | **0.5937** |
| | TEoUI | 6.9915 | **23.7730** | 3.4819 | 2.8488 | 0.5820 |
| | IBLA | 6.8470 | 31.4013 | 3.5331 | 1.4764 | 0.5918 |
| | ULAP | 6.7583 | 29.5713 | 3.4304 | 3.7060 | 0.5872 |
| Avg(Var) | | 6.77(0.09) | 29.82(9.99) | 3.68(0.16) | 2.24(1.9) | 0.56(0.001) |

## V. CONCLUSION AND DISCUSSION

Quality improvement methods of single underwater images based on image enhancement and color restoration are comprehensively reviewed to help researchers better explore this unknown underwater world. In this review, we firstly introduce the basic principles of underwater imaging model and selective light absorption characteristics under the water. Then we summarize the quality improvement methods of single underwater images into two categories: IFM-free underwater image enhancement and IFM-based underwater image restoration, thus describing existing methods and their characteristics. Finally, we provide an experimental-based comparison of the state-of-the-art quality improvement methods using multiple quality assessment metrics, which leads to discussing the issues confronted by the current IFM-free and IFM-based underwater image quality improvement methods. All in all, we provide a comprehensive outline of the progress and challenges of single underwater image quality improvement, which can help researchers in the further development of this area.

Although single underwater image enhancement and restoration methods have made tremendous progress, still today there is no algorithm that can be effectively applied to enhance underwater images captured from diverse environments, depths or scenes. The adaptability and robustness of underwater image enhancement methods still need to be improved.

In addition, traditional enhancement/restoration algorithms have relatively high complexity, which poses considerable limitations to our ability to scale up practical studies and applications.

The future works should focus on the follow aspects:

1) Improving the robustness and computational efficiency of underwater image enhancement methods. The desired image enhancement method should be able to adapt to various underwater conditions and develop an applicable enhancement strategy for different kinds of underwater applications. Through this review, we can find that none of the compared methods can improve the quality of all testing underwater images. IFM-based methods can recover actual scene but consume vast time to calculate two key optical parameters; whereas IFM-free methods can quickly enhance images by redistributing pixel values, but easily cause color distortion. A potential strategy for the quality improvement algorithms of underwater images is to wisely combine image restoration and enhancement. Meanwhile, many single image quality improvement methods without involving temporal coherence between adjacent frames cannot be directly employed in underwater video quality enhancement due to their excessive complexity.

2) Constructing a sufficient underwater image benchmark dataset. Until now, there is still a lack of publicly available underwater image datasets, including pairs of hazed and clear underwater images, underwater image background lights, and depth maps or transmission maps. IFM-based underwater image restoration methods require BL and TMs to recover real underwater scenes. The accuracy of BLs and TMs estimated by different methods and the effectiveness of these methods are compared and analyzed through subjective assessment due to the lack of a benchmark dataset. More deep learning techniques are applied in underwater image enhancement, e.g. using Generative Adversarial Networks (GAN) to regulate the white balance, and Recurrent Neural Networks (RNN) to de-noise and increase detail information. However, learning-based underwater image enhancement methods strongly depend on datasets, which requires a great number of the paired original and referenced images. Although synthetic datasets are often used to train deep models, there exists a big gap between the synthetic images and the actual underwater images obtained from complex underwater environments. Hence it is important to construct a public underwater image benchmark dataset with various pairs of hazed and enhanced images.

3) Establishing an effective underwater image quality assessment metric. A variety of image quality evaluation metrics are proposed; yet only few are suited to underwater images. In this review, the widely-used UCIQE and UIQM, which are inspired by the property of human vision system to quantify underwater color images, could not provide a fair assessment to the underwater image quality. Their evaluation favors the over-enhanced colorful images, which are against subjective preference to naturalness. Further research should be devoted to the smart combination of subjective and objective assessment and continuing to improve non-reference evaluation models.

4) Building a close relation between low-level image enhancement and high-level detection and classification. Current underwater image enhancement methods focus on improving the perceptual effect of images but ignore whether the enhanced images can increase the accuracy of high-level feature analysis such as target detection and classification. Hazed underwater images have similar objects with scene environment, which aggravates the difficulty of target recognition and detection. Thus, improving the quality of underwater images can effectively release the pressure of high-level underwater tasks. Therefore, in future studies, we can establish a high-level task, such as target detection under visibility degradation, and use the task completion as the criteria to evaluate an underwater image enhancement method.

5) Studying deep-sea image enhancement methods. Unlike shallow-water environment, the natural light (from the sun) propagating underwater will be fully absorbed under the sea below 1000 meters. Artificial light as the only light source has a strong influence on imaging. Intensity and point projection lighting limit vision range and cause uneven vignetting. The existing underwater image enhancement or restoration methods are not able to recover deep-sea images. Therefore, a new imaging model for deep-sea imaging environment is needed to solve light attenuation, uneven illumination, scattering interference and low brightness of deep-sea images, so as to improve the sense of reality of images and reduce the halo effect.